\documentclass[openacc]{rstransa}
\usepackage{units}

\begin{document}

\title{Proton Driven Plasma Wakefield Acceleration in AWAKE}


\author{
E. Gschwendtner$^{1}$, M. Turner$^{1}$, \\ **Author List Continues Next Page**}

\address{$^{1}$ CERN, Geneva, Switzerland\\}

\subject{Plasma Wakefield Acceleration, Proton Driven, Electron Acceleration}

\keywords{AWAKE, Plasma Wakefield Acceleration, Seeded Self Modulation}

\corres{Insert corresponding author name\\
\email{marlene.turner@cern.ch}}

\begin{abstract}
In this article, we briefly summarize the experiments performed during the first Run of the Advanced Wakefield Experiment, AWAKE, at CERN (European Organization for Nuclear Research). The final goal of AWAKE Run 1 (2013 - 2018) was to demonstrate that \unit[10-20]{MeV} electrons can be accelerated to GeV-energies in a plasma wakefield driven by a highly-relativistic self-modulated proton bunch. We describe the experiment, outline the measurement concept and present first results. Last, we outline our plans for the future.
\end{abstract}


\begin{fmtext}
\end{fmtext}
\maketitle 

\section*{Continued Author List}
E.~Adli$^{2}$,A.~Ahuja$^{1}$,O.~Apsimon$^{3,4}$,R.~Apsimon$^{3,4}$, A.-M.~Bachmann$^{1,5,6}$,F.~Batsch$^{1,5,6}$ \\ C.~Bracco$^{1}$,F.~Braunm{\"u}ller$^{5}$,S.~Burger$^{1}$,G.~Burt$^{7,4}$,
B.~Buttensch{\"o}n$^{8}$,A.~Caldwell$^{5}$,J.~Chappell$^{9}$, E.~Chevallay$^{1}$,M.~Chung$^{10}$,D.~Cooke$^{9}$,H.~Damerau$^{1}$, L.H.~Deubner$^{11}$,A.~Dexter$^{7,4}$,S.~Doebert$^{1}$, \\ J.~Farmer$^{12}$, V.N.~Fedosseev$^{1}$,R.~Fiorito$^{13,4}$,R.A.~Fonseca$^{14}$,L.~Garolfi$^{1}$,S.~Gessner$^{1}$, \\ B.~Goddard$^{1}$, I.~Gorgisyan$^{1}$,A.A.~Gorn$^{15,16}$,E.~Granados$^{1}$,O.~Grulke$^{8,17}$, 
A.~Hartin$^{9}$,A.~Helm$^{18}$, \\ J.R.~Henderson$^{7,4}$,M.~H{\"u}ther$^{5}$, M.~Ibison$^{13,4}$,S.~Jolly$^{9}$,F.~Keeble$^{9}$,M.D.~Kelisani$^{1}$,  S.-Y.~Kim$^{10}$, \\ F.~Kraus$^{11}$,M.~Krupa$^{1}$, T.~Lefevre$^{1}$,Y.~Li$^{3,4}$,S.~Liu$^{19}$,N.~Lopes$^{18}$,K.V.~Lotov$^{15,16}$, M.~Martyanov$^{5}$, \\ S.~Mazzoni$^{1}$,V.A.~Minakov$^{15,16}$, J.C.~Molendijk$^{1}$,J.T.~Moody$^{5}$,M.~Moreira$^{18,1}$,P.~Muggli$^{5,1}$, H.~Panuganti$^{1}$,A.~Pardons$^{1}$,F.~Pe\~na~Asmus$^{5,6}$, A.~Perera$^{13,4}$,A.~Petrenko$^{1,15}$,A.~Pukhov$^{12}$, S.~Rey$^{1}$,H.~Ruhl$^{21}$,H.~Saberi$^{1}$,A.~Sublet$^{1}$,P.~Sherwood$^{9}$,L.O.~Silva$^{18}$,A.P.~Sosedkin$^{15,16}$, P.V.~Tuev$^{15,16}$, F.~Velotti$^{1}$, L.~Verra$^{1,20}$,V.A.~Verzilov$^{19}$, J.~Vieira$^{18}$,C.P.~Welsch$^{13,4}$,M.~Wendt$^{1}$,B.~Williamson$^{3,4}$, M.~Wing$^{9}$, B.~Woolley$^{1}$,G.~Xia$^{3,4}$,  \\ The AWAKE Collaboration
\\ \\ $^{1}$ CERN, Geneva, Switzerland \\ $^{2}$ University of Oslo, Oslo, Norway \\ $^{3}$ University of Manchester, Manchester, UK \\  $^{4}$ Cockcroft Institute, Daresbury, UK \\ $^{5}$ Max Planck Institute for Physics, Munich, Germany \\ $^{6}$ Technical University Munich, Munich, Germany  \\$^{7}$ Lancaster University, Lancaster, UK \\ $^{8}$ Max Planck Institute for Plasma Physics, Greifswald, Germany \\$^{9}$ UCL, London, UK \\ $^{10}$ UNIST, Ulsan, Republic of Korea \\ $^{11}$ Philipps-Universit{\"a}t Marburg, Marburg, Germany \\ $^{12}$ Heinrich-Heine-University of D{\"u}sseldorf, D{\"u}sseldorf, Germany \\ $^{13}$ University of Liverpool, Liverpool, UK \\ $^{14}$ ISCTE - Instituto Universit\'{e}ario de Lisboa, Portugal \\ $^{15}$ Budker Institute of Nuclear Physics SB RAS, Novosibirsk, Russia \\ $^{16}$ Novosibirsk State University, Novosibirsk, Russia \\ $^{17}$ Technical University of Denmark, Lyngby, Denmark \\ $^{18}$ GoLP/Instituto de Plasmas e Fus\~{a}o Nuclear, Instituto Superior T\'{e}cnico, Universidade de Lisboa, Lisbon, Portugal \\ $^{19}$ TRIUMF, Vancouver, Canada \\ $^{20}$ University of Milan, Milan, Italy \\ $^{21}$ Ludwig-Maximilian University

\newpage
\section{Introduction}
Charged particle acceleration in conventional radio-frequency cavities is limited to gradients on the order of \unit[100]{MeV/m}, due to electrical breakdown in the metallic structures. Plasma wakefield acceleration can exceed this limit and offers the possibility to accelerate charged particles with gradients on the order of GeV/m or higher. The maximum accelerating gradient can be estimated from the non-relativistic, cold plasma wavebreaking field ($E_{WB}$):
\begin{equation}
e E_{WB} = m_e c \omega_{pe} \sim 100\frac{eV}{m}\cdot \sqrt{n_{pe}[\text{cm}^{-3}]},
\label{eq:WB}
\end{equation}
with the electron mass $m_e$, the plasma electron frequency $\omega_{pe}=\sqrt{n_{pe} e^2/(m_e \epsilon_0)}$,  the plasma electron density $n_{pe}$, the electron charge $e$, the vacuum dielectric constant $\epsilon_0$ and the speed of light $c$. Equation \ref{eq:WB} shows that in order to reach GeV/m gradients, the plasma density needs to exceed \unit[$n_{pe}>10^{14}$]{electrons/cm$^{3}$}.


Plasma wakefields are typically excited either by a relativistic particle bunch or an intense laser pulse. The distance over which the bunch or pulse can drive high gradient wakefields depends on its stored energy. Available laser pulses and electron bunches carry energies of around \unit[$20$]{J}, but proton bunches (for example at the European Organization for Nuclear Research, CERN) can carry energies of many tens of kilo-Joules. From the energy conservation point of view, this is enough energy to accelerate \unit[1]{nC} of electrons to hundreds of TeV (we note that the transfer efficiency in this scheme will be much lower than \unit[100]{\%}). 



In 2009, A. Caldwell et al. \cite{ALLEN} proposed to accelerate witness electrons in plasma wakefields driven by a Large Hadron Collider (LHC) type proton bunch. In their simulations, they use a \unit[1]{TeV} proton bunch (with \unit[$10^{11}$]{protons/bunch} and thus an energy of \unit[16]{kJ}) with an rms longitudinal size of \unit[0.1]{mm} and drive wakefields over a distance of \unit[600]{m}. Witness electrons reach an energy of \unit[0.6]{TeV}.

To effectively excite wakefields, the rms length ($\sigma_z$) and transverse size ($\sigma_r$) of the Gaussian drive bunch should be matched to the plasma density: $\sigma_z\approx\sqrt{2}c/\omega_{pe}$ and  $\sigma_r\approx c/\omega_{pe}$ (\unit[$c/\omega_{pe}<0.53$]{mm} for \unit[$n_{pe}>10^{14}$]{electrons/cm$^3$}). While available high-energy proton bunches can be focused tightly (\unit[$\sigma_r<0.2$]{mm}), their rms bunch length $\sigma_z$ is on the order of \unit[10]{cm} and thus much too long to excite high amplitude wakefields. To still use those bunches, one can rely on the self-modulation instability \cite{PoP2-1326,PoP4-1154,SMI1,SMI2} (or if seeded, the seeded self-modulation (SSM) \cite{SSM}) to modulate a long proton bunch into a train of micro-bunches. After the modulation process saturates, microbunches have a length on the order of $\sigma_z\approx\sqrt{2}c/\omega_{pe}$ and are spaced at the plasma electron wavelength $\lambda_{pe}=2\pi c/\omega_{pe}$. This bunch train can then resonantly excite high-amplitude wakefields in a uniform plasma.

In this article, we summarize the experimental efforts and achievements of proton driven plasma wakefield acceleration in the context of the Advanced WAKefield Experiment (AWAKE) \cite{AWAKE} at CERN (see Sec. \ref{sec:AWAKE}). AWAKE ran experiments from December 2016 to November 2018. The experimental program focused on two main topics: first, the demonstration of the seeded self-modulation of a long proton bunch in plasma (Run 1, phase 1: 2016, 2017) \cite{SSM}, see Sec. \ref{sec:SSM}; second, the demonstration of the acceleration of electrons in a plasma wave that was resonantly excited by the self-modulated proton bunch (Run 1, phase 2: 2018), see Sec. \ref{sec:electronacc}. We additionally outline plans and goals for the short term (Run 2: 2021-2024) and long-term ($>$2026) future in Sec. \ref{sec:future}.

\section{The AWAKE Experiment at CERN}
\label{sec:AWAKE}

\begin{figure}[htb!]
    \centering
    \includegraphics[width=\textwidth]{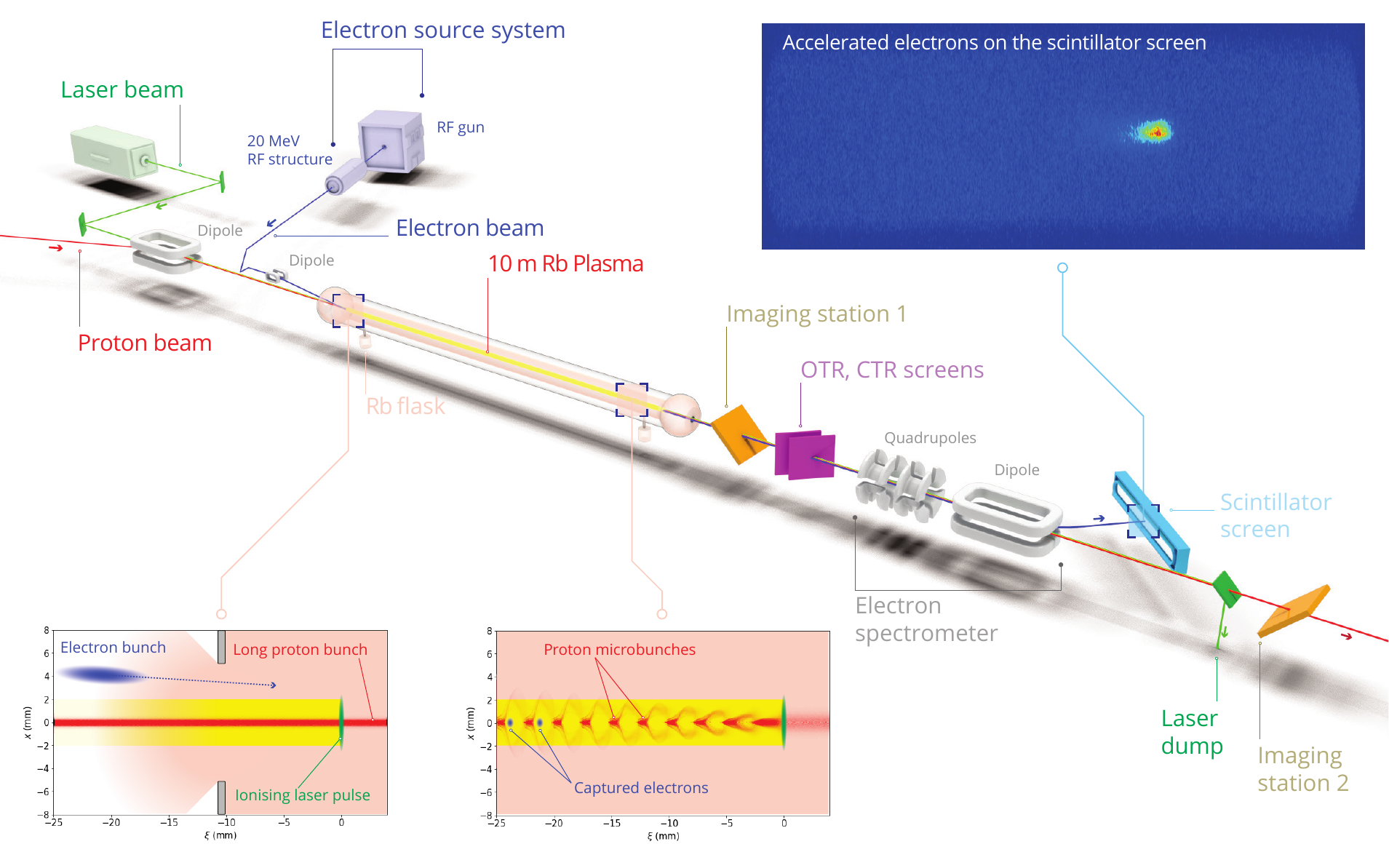}
    \caption{Schematic layout and description of the AWAKE experimental facility, beams and diagnostics. The insert panel on the bottom left shows a schematic of the spatial and temporal alignment of the proton, laser and electron bunch entering the vapor source; the bottom middle panel shows a schematic of the transverse and longitudinal proton microbunch density structure in plasma (after the self-modulation process saturated); the panel on the top right shows an experimental image obtained by the spectrometer camera, for when electrons were accelerated.}
    \label{fig:schematic}
\end{figure}

In AWAKE, a proton bunch drives plasma wakefields. The bunch consists of \unit[$3\times10^{11}$]{protons} with a particle momentum of \unit[400]{GeV/c} (\unit[19.2]{kJ}). The bunch is focused at the plasma entrance to a transverse rms size of \unit[$\sigma_r\approx 0.2$]{mm}. Its rms bunch length can be adjusted from 6 to \unit[12]{cm}. Figure~\ref{fig:schematic} shows a schematic drawing of the AWAKE experimental facility.

The AWAKE plasma is \unit[10]{m} long, has a radius of approximately \unit[$1$]{mm} and a density uniformity better than \unit[0.1]{\%} \cite{VaporSource} (to allow resonant wakefield excitation from hundreds of microbunches). To create the plasma, we evaporate rubidium in a heat exchanger  and ionize the outermost electron of each rubidium atom with a laser pulse (pulse length: \unit[100]{fs}, pulse energy: \unit[$\leq450$]{mJ}). The vapor (and thus also the plasma) density is controlled by the temperature of the source and is adjustable between \unit[$0.5-10\times10^{14}$]{atoms/cm$^3$}. To keep the density uniform within \unit[0.1]{\%}, the temperature along the source is constant within \unit[0.05]{K}. 

Since there is no window that allows a proton, electron and laser pulse to enter the vapor source at the same time, the system is open at both ends. Rubidium atoms flow out of the opening apertures (the aperture radius is \unit[5]{mm}) and condensate on the walls of the cold expansion volumes (kept below the rubidium freezing point of \unit[39]{$^{o}$C}). The length of the density ramps (at both entrance and exit of the plasma) is on the order of the opening aperture \cite{DensityRamp}.

\begin{figure}[htb!]
    \centering
    \includegraphics[width=0.8\textwidth]{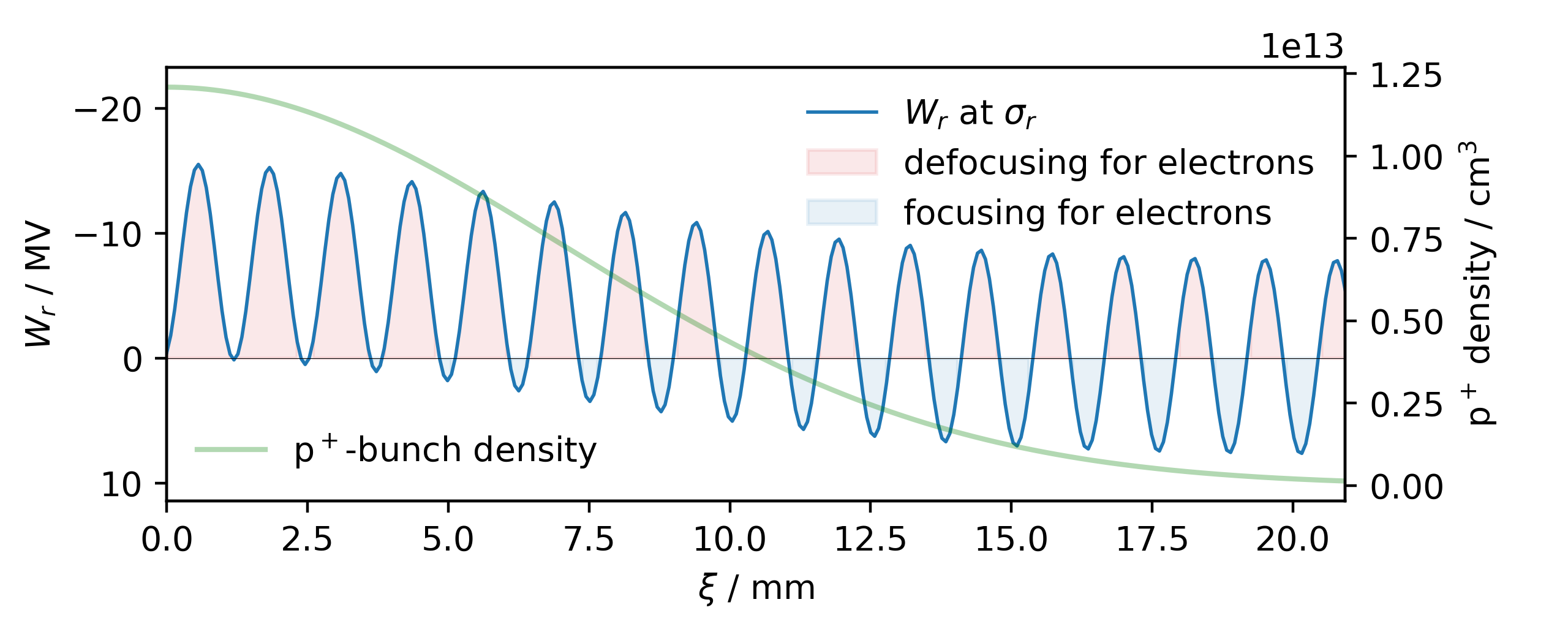}
    \caption{Transverse ($W_r$ at $\sigma_r$) initial seed wakefields along the proton bunch ($\xi$) in plasma with an electron density of \unit[$7\times10^{14}$]{cm$^{-3}$}. The proton bunch rms transverse size is \unit[$\sigma_r =0.15$]{mm}. To speed up the calculation, the bunch length was made 10 times shorter (\unit[$\sigma_z=7$]{mm} instead of \unit[$\approx$7]{cm}) and the bunch intensity 10 times less (\unit[$3\times10^{10}$]{protons/bunch} instead of \unit[$3\times10^{11}$]{protons/bunch}). Since the calculation is linear, the resulting wakefield amplitudes are the same. The seeding position ($\xi=0$) is in the center of the proton bunch. The green line shows the particle density along the bunch ($\xi$). Red areas mark wakefield phase regions that are defocusing for electrons, blue areas marks focusing regions. The bunch propagates to the left.}
    \label{fig:seedfields}
\end{figure}

The laser pulse is additionally used to seed the self-modulation process. In principle, the self-modulation of a long proton bunch can grow from the random noise in a preformed plasma. Then, the wakefields phase and amplitude varies from event to event and controlled electron injection (into the accelerating, focusing phase) is not possible. To fix the wakefields phase and amplitude, we seed the self-modulation process: we overlap the laser pulse in space and time with the proton bunch. The laser pulse creates an ionization front (see bottom panels on Fig. \ref{fig:schematic}) inside the proton bunch. Protons ahead of the laser pulse propagate in rubidium vapor, the ones after in the plasma (the transition is sharp due to the short length of the laser pulse). Figure \ref{fig:seedfields} shows the initial transverse and longitudinal seed fields along the proton bunch at the plasma entrance for when it is seeded sharply at the center. We note that this initial wakefield amplitudes ($W_{r}$) are on the order of \unit[10]{MV/m} much larger than the expected noise level of \unit[10]{kV/m} \cite{LotovSeeding}. During the self-modulation process the wakefield amplitude grows from the initial seed value according to the self-modulations growth rate until saturation.  

To probe the plasma wakefields and to demonstrate electron acceleration, we externally inject \unit[10-20]{MeV} electrons. A  \unit[$\sim5.2$]{ps} long UV laser pulse produces \unit[100-600]{pC} of electrons from a Cesium-Tellurium (Cs2Te) cathode, which are accelerated  to \unit[$\sim5$]{MeV} in a 2.5 cell radio-frequency photo injector \cite{ElectronSource}. To facilitate the timing requirements of the experiment, the electron bunch length ($\sigma_z>$ \unit[4]{ps}) is longer or on the order of the plasma electron wavelength ($\lambda_{pe}<$ \unit[8]{ps}).
The bunch is then further accelerated to \unit[$10-20$]{MeV} by a one meter long S-Band booster structure, and transported to the plasma entrance \cite{ElectronTransfer}. 
A quadrupole triplet focuses the electron bunch to a transverse rms size of \unit[0.2-2]{mm} from approximately \unit[$3$]{m} upstream to approximately \unit[$5$]{m} downstream the plasma entrance (the further the focal point from the triplet, the larger the transverse size at focus).

\section{Seeded Self-Modulation}
\label{sec:SSM}
To experimentally observe and study the seeded self-modulation of a long ($\sigma_z\gg\lambda_{pe}$) proton bunch in plasma we align the laser pulse to the center of the proton bunch (in space and time). When there was either no laser pulse (or the laser was behind the proton bunch) or no rubidium, the bunch propagated according to its bunch and optics parameters. When the laser pulse was seeding we observed radial proton bunch self-modulation. 

The focusing regions of the plasma wakefields form a microbunch train. In \cite{RiegerSSM} we show that the frequency of the microbunches corresponds to the plasma electron frequency over the measured density range. In between the focusing regions, we observed defocused protons. Defocused protons appeared at large radial positions ($r\gg\sigma_r$). In \cite{TurnerSSM} we prove that the wakefield amplitudes grow along the bunch and the plasma significantly evolves from the initial seed fields shown in Fig. \ref{fig:seedfields}, due to the seeded self-modulation process.

Additionally we studied:
\begin{itemize}
    \item proton bunch defocusing and wakefield growth as a function of plasma electron densities, proton bunch populations, plasma electron density gradients and laser pulse seed locations along the proton bunch; 
    \item competition between self-modulation and the hose instability for different proton bunch populations, proton bunch shapes and laser pulse seed locations;
    \item phase stability of the wakefields and the minimum seed level to observe it;
    \item reproducibility of the self-modulation instability (unseeded), the laser pulse seeded self-modulation and the electron bunch seeded self-modulation.
\end{itemize}

The results and details of these additional studies will be submitted for publication in the near future.

\section{Electron Acceleration}
\label{sec:electronacc}
To prove that electrons can be accelerated in plasma wakefields driven by a self-modulating proton bunch, we inject \unit[$\sim18$]{MeV} electrons close to the plasma entrance. This injection was challenging because of two main reasons: first, the initial seed fields (driven by the proton bunch) are largely defocusing for electrons (see red areas marked in Fig.~\ref{fig:seedfields}). Their amplitude (\unit[$\approx$10]{MV/m}) is large enough to defocus the \unit[$\approx20$]{MeV} electrons within short distances (electrons gain angles of \unit[$\sim50$]{mrad} over \unit[10]{cm} without acceleration). The seed wakefields include electron focusing regions only around one rms length $\sigma_z$ behind the seed point (as indicated by blue regions in Fig. \ref{fig:seedfields}) or after the proton bunch density starts to modulate. Second, the wakefields phase is evolving as 1) the plasma electron wavelength is changing along the density ramp and 2) the proton bunch density is modulating over the first few meters of plasma (see Sec. \ref{sec:AWAKE}). 

Thus, we aimed to inject electrons into the wakefields, once their phase does not evolve strongly anymore. The concept and idea behind the electron injection scheme is discussed in \cite{AAC2018}. Accelerated electrons were detected on an imaging magnetic spectrometer \cite{Spectro}.

\begin{figure}[htb!]
    \centering
    \includegraphics[width=0.8\textwidth]{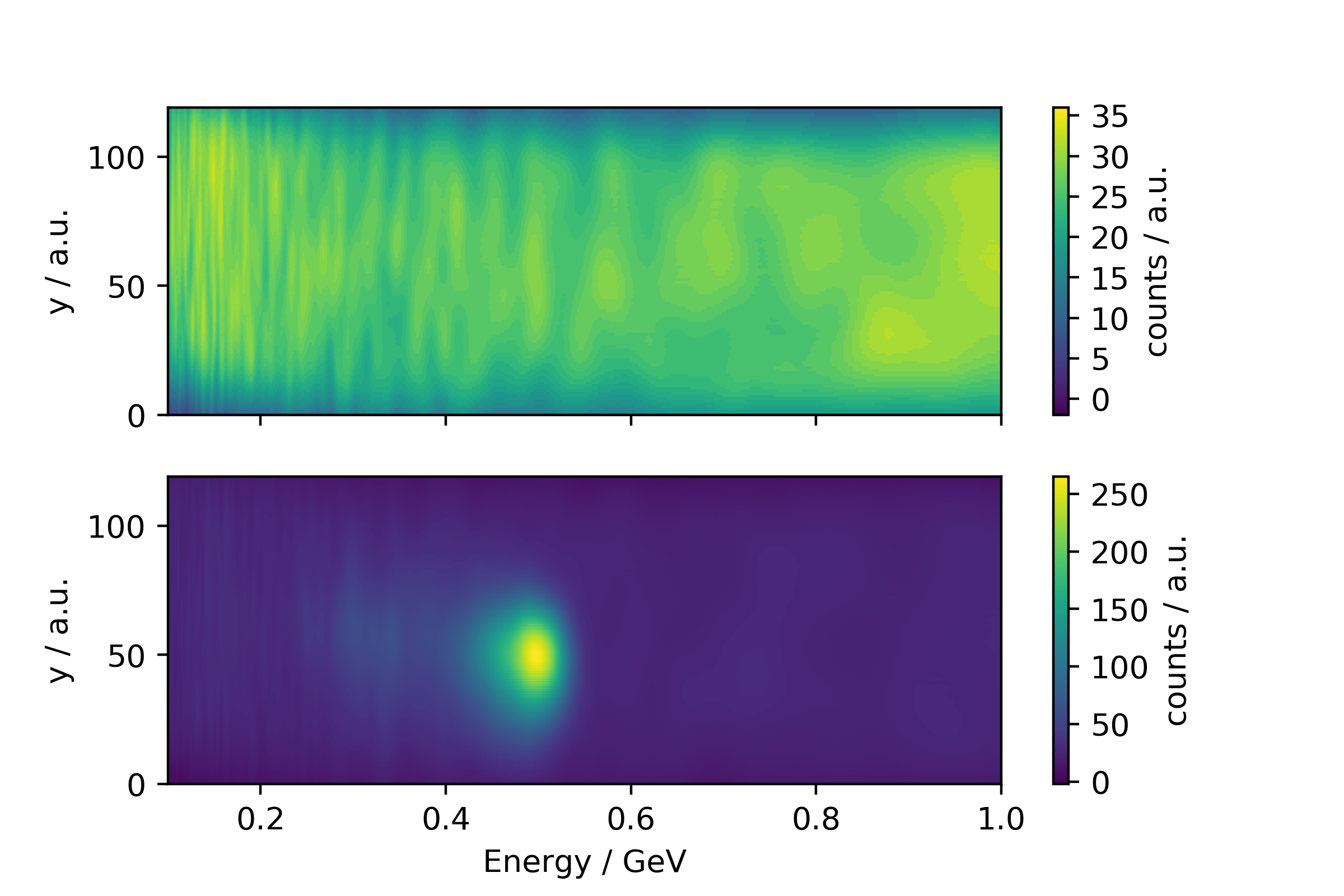}
    \caption{Top panel: spectrometer background image; lower panel: example of a spectrometer image for when electrons were accelerated. Note the change on the maximum number of counts between the two images. The same constant background was subtracted from each image.}
    \label{fig:acconoff}
\end{figure}

Figure \ref{fig:acconoff} compares a spectrometer background image (top panel) to an image where accelerated electrons were observed (lower panel). The electron signal is clearly visible above the background. We note that the accelerated electrons energy spread is peaked and finite. Detailed studies on the accelerated electron bunch quality and stability will be published elsewhere. 

\begin{figure}[htb!]
    \centering
    \includegraphics[width=0.8\textwidth]{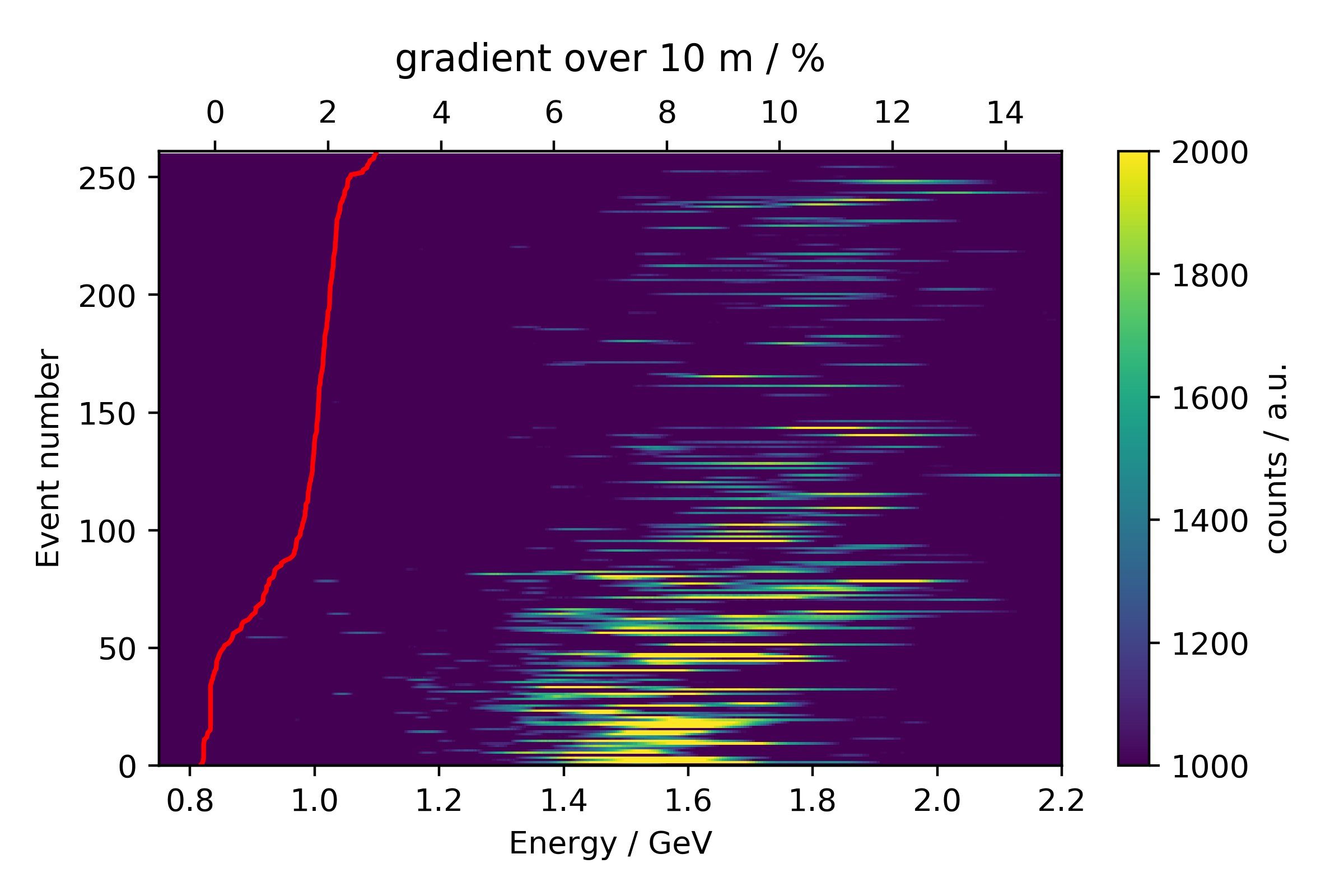}
    \caption{Waterfall plot of vertically summed spectrometer images during a change of plasma electrons density gradient. The upstream density is \unit[$7\times10^{14}$]{cm$^{-3}$}. The red line indicates the measured density gradient for each event. Note that integrated signals below 1000 counts are cut from the image.}
    \label{fig:7e14acc}
\end{figure}

The first acceleration results were recently published in \cite{ElectronAcc}. In the article, we present the maximum electron energy observed for plasma densities of \unit[1.8, 3.9 and 6.6$\times 10^{14}$]{electrons/cm$^{3}$} (with and without plasma electron density gradients). We show that electrons were accelerated up to energies of \unit[2]{GeV}.

Figure  \ref{fig:7e14acc} shows a waterfall-plot of more than 250 consecutive events (with the same current strengths in the spectrometer system) sorted by plasma electron density gradient (red line in Fig. \ref{fig:7e14acc}) for a plasma electron density (at the plasma entrance) of \unit[$6.6\times10^{14}$]{cm$^{-3}$}. As discussed by the AWAKE collaboration in \cite{ElectronAcc}, the energy of the accelerated electrons can increase with a positive gradient, under certain circumstances.
As simulations show \cite{gradient}, a positive plasma electron density gradient can compensate some of the phase shift that occurs during the self-modulation process. Electrons that would undergo cycles of acceleration and deceleration in a uniform plasma (due to the wakefields phase shift) stay in the accelerating wakefield phase and thus gain more energy.

We note that no cuts are applied to the data presented in Fig. \ref{fig:7e14acc}, even though some experimental parameters fluctuate (e.g. proton bunch charge, electron beam position...).

Furthermore, we probed the electron acceleration and studied the structure of the accelerating wakefield as a function of:
\begin{itemize}
    \item injection angle and injection location (along the plasma);
    \item plasma electron density, plasma electron density gradient and proton bunch population;
    \item time-delay between the seeding laser pulse and electron bunch for different seed locations in the proton bunch.
\end{itemize}

Analysis of the results is ongoing.

\section{Future plans}
\label{sec:future}
AWAKE successfully finished the main experimental program of Run 1. It was demonstrated for the first time that a long proton bunch self-modulates over \unit[10]{m} of plasma and that electrons can be accelerated in its wake to 2~GeV. In a next step, AWAKE Run 2 (2021 - 2024) aims to demonstrate that one can accelerate an electron bunch to high energies, about 10~GeV, while maintaining its bunch quality and to show that the concept is scalable to long acceleration distance scales. 

The idea for Run 2 is to split the plasma into two sections: a self-modulator and an accelerator. The proton bunch modulates (until saturation is reached) into a micro-bunch train in the self-modulator. A short electron bunch ($\sigma \sim$ \unit[100]{fs}) is injected into the stable wakefields (driven by the microbunch train) in the accelerator plasma. First simulation results \cite{Veronica} indicate that the majority of the witness bunch can preserve its quality (emittance, relative energy spread) if: 1) the electron bunch beam loads the wakefields and 2) its front contributes to creating a blow-out region in which its back can experience a linear focusing force. 

The final goal by the end of AWAKE Run 2 is to be in a position to use the AWAKE scheme for first high-energy physics applications such as fixed target experiments for dark photon searches and future electron-proton or electron-ion colliders~\cite{VHEP}. First studies show the feasibility of a fixed target experiment in the AWAKE facility: the expected number of electrons on target exceeds the current number of the NA64 \cite{NA64} experiment by four orders of magnitude, making it a very promising option in the search for new physics.

Studies show that electrons can be accelerated to 70 GeV in a 130 m long plasma installed in an extended extraction tunnel from CERN Super Proton Synchrotron (SPS) to the LHC and transported to collision with protons/ions from the LHC. The experiment would focus on studies of the structure of matter and QCD in a new kinematic
domain.

\section{Conclusion and Summary}
We introduced the motivation and concept of proton-driven plasma wakefield acceleration and explained the AWAKE experiment. In AWAKE Run 1 we demonstrated that: a proton bunch ($\sim$two orders longer than the plasma electron wavelength $\lambda_{pe}$) self-modulates over \unit[10]{m} of plasma; that this self-modulated bunch train can resonantly excite high-amplitude plasma wakefields and that electrons can be accelerated in the resonantly driven wakefield. The goal for future experiments is to accelerate an electron bunch with good beam quality to high energies and to apply the scheme  for new physics studies.

\section*{Acknowledgements}
The support of the Max Planck Society is gratefully acknowledged. This work was supported in parts by the Siberian Branch of the Russian Academy of Science (project No. 0305-2017-0021), a Leverhulme Trust Research Project Grant RPG-2017-143 and by STFC (AWAKE-UK, Cockroft Institute core and UCL consolidated grants), United Kingdom; a Deutsche Forschungsgemeinschaft project grant PU 213-6/1 ``Three-dimensional quasi-static simulations of beam self-modulation for plasma wakefield acceleration''; the National Research Foundation of Korea (Nos.\ NRF-2015R1D1A1A01061074 and NRF-2016R1A5A1013277); the Portuguese FCT---Foundation for Science and Technology, through grants CERN/FIS-TEC/0032/2017, PTDC-FIS-PLA-2940-2014, UID/FIS/50010/2013 and SFRH/IF/01635/2015; NSERC and CNRC for TRIUMF's contribution; and the Research Council of Norway. M. Wing acknowledges the support of the Alexander von Humboldt Stiftung and DESY, Hamburg. The AWAKE collaboration acknowledge the SPS team for their excellent proton delivery.


\begin{thebibliography}{99}
\bibitem{ALLEN} A. Caldwell et al., Proton-driven plasma-wakefield acceleration. Nature Physics 5, 363-367  (2009).

\bibitem{PoP2-1326} J. Krall and G. Joyce, Transverse equilibrium and stability of the primary beam in the plasma wake-field accelerator. Phys. Plasmas 2, 1326 (1995).
 
\bibitem{PoP4-1154} D.H. Whittum, Transverse two-stream instability of a beam with a Bennett profile. Phys. Plasmas 4, 1154 (1997).

\bibitem{SMI1} N. Kumar, et al., Self-modulation instability of a long proton bunch in plasmas, PRL 104 (25), 255003 (2010).

\bibitem{SMI2} A. Pukhov, et al., Phase velocity and particle injection in a self-modulated proton-driven plasma wakefield accelerator, PRL 107 (14), 145003 (2011).

\bibitem{SSM} P. Muggli and the AWAKE Collaboration, AWAKE readiness for the study of the seeded self-modulation of a 400 GeV proton bunch. Plasma Phys. Control. Fusion 60, 1, (2017).

\bibitem{AWAKE} E. Gschwendtner and the AWAKE Collaboration, AWAKE, The Advanced Proton Driven Plasma Wakefield Acceleration Experiment at CERN. Nucl. Instrum. Methods Phys. Res. A 826 76-82 (2016).

\bibitem{VaporSource} E. Oz et al., A novel Rb vapor plasma source for plasma wakefield accelerators. Nucl. Instrum. Methods Phys. Res. A 740, 197-202, (2014).

\bibitem{DensityRamp} G. Plyushchev, A Rubidium Vapor Source for a Plasma Source for AWAKE. J. Phys. D 51 025203 (2017).

\bibitem{LotovSeeding} K.V. Lotov et al., Natural noise and external wakefield seeding in a proton-driven plasma accelerator. Phys. Rev. AB 19 041301 (2013).

\bibitem{ElectronSource} K. Pepitone et al., The electron accelerators for the AWAKE experiment at CERN baseline and future developments, Nucl. Instrum. Methods Phys. Res., Sect. A, in press.

\bibitem{ElectronTransfer} J.S. Schmidt et al., The AWAKE electron primary beamline,  Proceedings of the 6th International Particle Accelerator Conference, pp. 2584-2586, (2015).

\bibitem{RiegerSSM} The AWAKE Collaboration, Experimental Observation of Proton Bunch Modulation in a Plasma at Varying Plasma Densities, PRL 122,054802 (2019).

\bibitem{TurnerSSM} M.Turner and the AWAKE Collaboration, Experimental Observation of Plasma Wakefield Growth Driven by the Seeded Self-Modulation of a Proton Bunch, PRL 122,054801 (2019).

\bibitem{AAC2018} M. Turner et al., External Electron Injection for the AWAKE Experiment, Proceedings of AAC 2018 Breckenridge, IEEE, p. 195.

\bibitem{Spectro} F. Keeble et al., The AWAKE Electron Spectrometer, in Proc. 9th Int. Particle Accelerator Conf. (IPAC'18), Vancouver, Canada, Apr.-May 2018, pp. 4947-4950. doi:10.18429/JACoW-IPAC2018-THPML118.

\bibitem{ElectronAcc} AWAKE Collaboration, Acceleration of electrons in the plasma wakefield of a proton bunch, Nature 561, 363 (2018).

\bibitem{gradient} A. Petrenko et al., Numerical Studies of Electron Acceleration behind Self-Modulating Proton Beam in Plasma with a Density Gradient, NIM A 829, pp. 63-66 (2016).

\bibitem{Veronica} V. Olsen et al., Emittance preservation of an electron beam in a loaded quasilinear plasma wakefield, Phys. Rev. Accel. Beams 21, 011301 (2018).

\bibitem{VHEP} A. Caldwell and M. Wing, VHEeP: a very high energy electron proton collider, Eur. Phys. J C 76, 463 (2016).

\bibitem{NA64} na64.web.cern.ch

\end{thebibliography}
\end{document}